\documentclass[twocolumn]{webofc}
\usepackage[varg]{txfonts}
\usepackage{lipsum}
\usepackage{siunitx}
\usepackage{acronym}
\DeclareSIUnit\year{yr}
\acrodef{Auger}{The Pierre Auger Observatory}
\acrodef{UHECR}{Ultra-High-Energy Cosmic Rays}
\acrodef{SSD}{Scintillator Surface Detector}
\acrodef{EAS}{Extensive Air Shower}
\acrodef{WLS}{Wave-Length Shifting}
\acrodef{PMT}{Photomultiplier Tube}
\acrodef{XPS}{Extruded Polystyrene}
\acrodef{EPS}{Expanded Polystyrene}
\acrodef{PE}{Photoelectron}
\acrodef{WCD}{Water-Cherenkov Detector}
\acrodef{MIP}{Minimum Ionizing Particle}
\acrodef{VMIP}{Vertical Minimum Ionizing Particle}
\acrodef{SPE}{Single Photoelectron}
\acrodef{SD}{Surface Detector}
\acrodef{EA}{Engineering Array}
\def\Offline{\mbox{$\overline{\text{Off}}$\hspace{.05em}\protect
\raisebox{.4ex}{$\protect\underline{\text{line}}$}}\xspace}
\begin{document}
\title{Scintillator Surface Detector simulations for AugerPrime}

\author{\firstname{David} \lastname{Schmidt}\inst{1}\fnsep
\thanks{\email{david.schmidt@kit.edu}}
\lastname{for the Pierre Auger Collaboration}\inst{2}\fnsep
\thanks{\email{auger_spokespersons@fnal.gov}}\fnsep
\thanks{Authors list: http://www.auger.org/archive/authors\_2018\_10.html}}

\institute{
  Karlsruhe Institute of Technology (KIT)
  \and
  Observatorio Pierre Auger, Av. San Martín Norte 304, 5613 Malargüe, Argentina
}

\abstract{
Knowledge of the mass composition of ultra-high-energy cosmic rays is
understood to be a salient component in answering the open questions in the
field. The AugerPrime upgrade of the Pierre Auger Observatory aims to enhance
its surface detector with the hardware necessary to reconstruct primary mass
for individual events. This involves placing a scintillation-based detector
with an active area of \SI{3.8}{\meter\squared} on top of each existing
water-Cherenkov detector in its surface detector array. Here, we present the
methods for simulating this Scintillator Surface Detector. These simulations
have and will continue to aid in the interpretation of measurements with
AugerPrime as well as the development and improvement of event reconstruction
algorithms including primary mass.
}
\maketitle
\section{Introduction}
\label{s:introduction}

\ac{Auger} \cite{Auger}, with its unprecedented exposure of over \SI{60000}{
\kilo\meter\squared\steradian\year} acquired during more than decade of data
collection, has led to progress in the field of \ac{UHECR} physics in a number
of ways. To name a couple, the suppression in the energy spectrum above
approximately \SI{40}{EeV} has been confirmed to high precision \cite{Cutoff1,
Cutoff2}, and a large-scale dipole anisotropy above \SI{8}{EeV} has been
clearly observed \cite{Dipole}. With these advances, however, \ac{Auger} has
also uncovered additional and unexpected complexity. Perhaps most notable is
the trend towards heavier composition at the highest energies, which reopens
questions regarding the origins of the flux suppression. Additionally, the
long standing mystery as to the origins of \acp{UHECR} remains unsolved, and
the additional complication of mass now presents further challenges. For this
reason, the observatory is embarking on its next phase by equipping its \ac{SD}
array with the hardware necessary to estimate the mass of \acp{UHECR} on an
event-by-event basis.

This upgrade, known as AugerPrime \cite{AugerPrime, AugerPrimeICRC,
AugerPrimeUHECR}, capitalizes on the fact that the magnitude of the muonic
component of \acp{EAS} produced by \acp{UHECR} scales with primary mass.
AugerPrime aims to disentangle the contributions of the electromagnetic and
muonic shower components in \ac{SD} measurements, thereby obtaining a 
mass-sensitive estimator. This is being accomplished through the installation
of an \ac{SSD} on top of each of the existing \acp{WCD} that make up the
\ac{SD} array. The relatively higher sensitivity of the \acp{SSD} (\acp{WCD})
to particles of the electromagnetic (muonic) air shower components allows for
this disentanglement.

A prototype array of \acp{SSD} has been operating in the field since September
of 2016, and deployment of production design \acp{SSD} for the complete array
is currently underway. In order to aid in understanding and interpreting
measurements of the upgraded \ac{SD}, as well as develop reconstruction
algorithms, we have developed comprehensive \ac{SSD} simulations. The methods
used in these simulations are described in this proceeding and build upon
\cite{AugerPrimeOffline}.

\section{Scintillator Surface Detector}
\label{s:ssd}
\begin{figure}[h]
\centering
\includegraphics[width=0.45\textwidth,clip]{{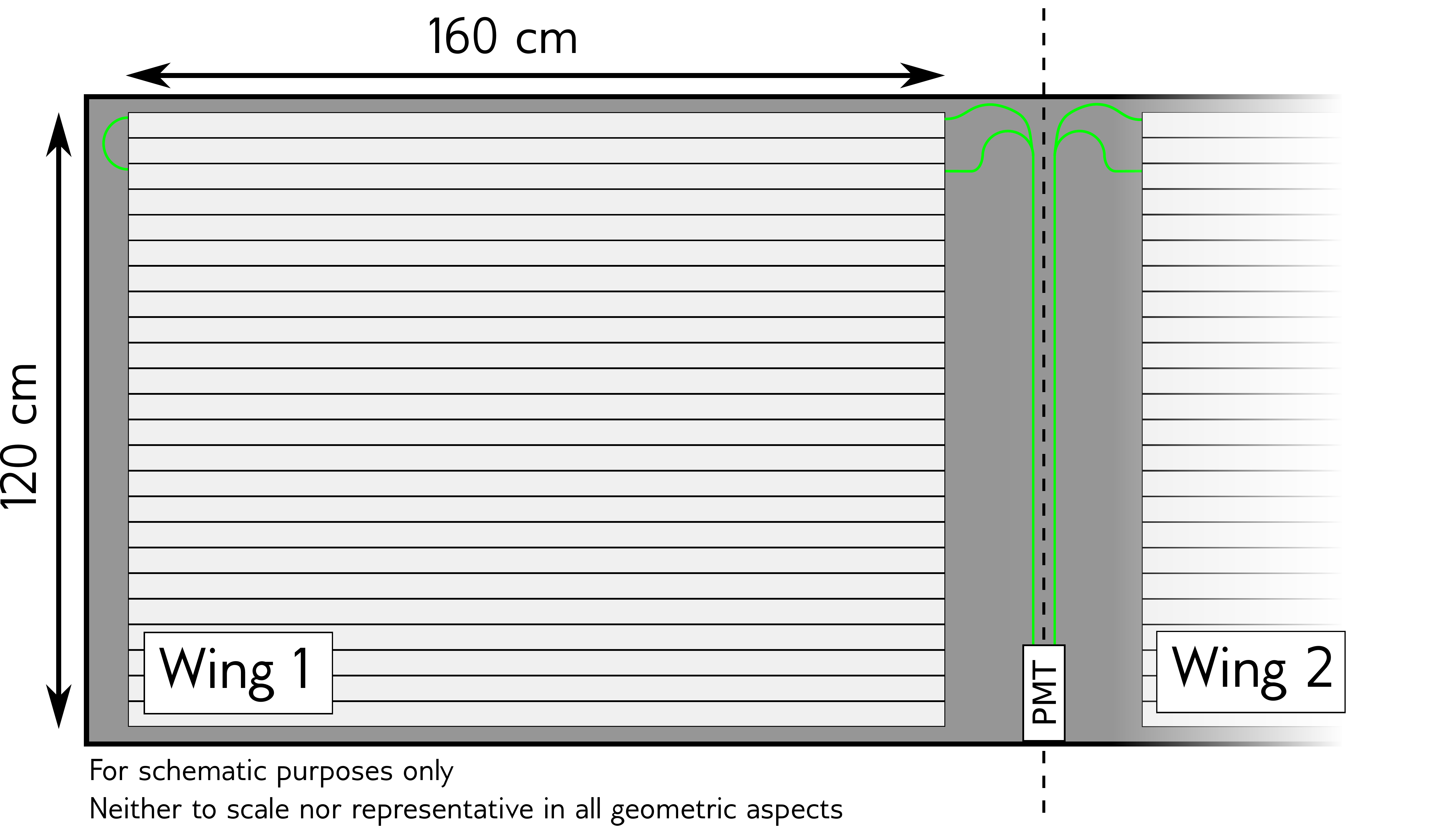}}
\caption{Top-down schematic depiction of the arrangement of bars and fibers
within the \ac{SSD} casing. The written dimensions describe the active
surfaces of the detectors, namely the scintillator bars.}
\label{f:geo_schem}
\end{figure}

The active area of an \ac{SSD} is comprised of 48 polystyrene scintillator
bars measuring \SI{1600}{\milli\meter} by \SI{50}{\milli\meter} with a
thickness of \SI{10}{\milli\meter}. This makes for an active area of just over
\SI{3.8}{m^2}, which is split between two wings of 24 bars each
(see Fig. \ref{f:geo_schem}). \SI{1}{\milli\meter} diameter \ac{WLS} fibers of
\SI{5.8}{\meter} length are routed through two bean shaped holes inside each
bar with a U-bend of \SI{10}{\centi\meter} diameter at the ends furthest from
the central axis of the \ac{SSD} such that both fiber ends may be routed into
a cookie situated between the two scintillator wings. There are approximately
\SI{1.25}{\meter} of fiber between the central ends of the scintillator bars
and the cookie, and the lengths of the fibers are uniform for all scintillator
bars through which they are routed. The bars and fibers are housed within an
aluminum casing whose bottom consists of a composite panel with
\SI{22}{\milli\meter} of \ac{XPS} situated between two \SI{1}{\milli\meter}
sheets of aluminum. The top of the housing consists of a \SI{1}{\milli\meter}
sheet of aluminum. The space within the \ac{SSD} housing not occupied by the
bars and fibers is mostly filled with \ac{EPS}, resulting in a volume of air
inside the detector of less than \SI{10}{L}. For a detailed description of the
detector materials, geometry, and construction, see \cite{SSDICRC2017}.
\section{Simulation Application}
\label{s:simulations}
The \ac{SSD} simulation application is housed within Auger's official
simulation and reconstruction framework, \Offline \cite{Offline}, which was
upgraded to accommodate for the demands of the large-scale detector upgrade
that is AugerPrime \cite{AugerPrimeOffline}. The \ac{SSD} simulations are
based in the field-standard simulation software Geant4 \cite{Geant4} (version
4.10), but, where it is possible, make use of measurements of \acp{SSD}
performed with a centimeter-precision muon telescope for enrichment and
tuning. The combined use of Geant4 and the muon telescope measurements has
made highly detailed simulation of the detector possible. The simulations
account for the energy-loss processes of traversing particles, the decays
occurring from excited states of atoms within the scintillator bars and
\ac{WLS} fibers, the attenuation of photons along the fibers, the \acp{PE}
generated at the photo-cathode of the \ac{PMT}, the current at the base of the
\ac{PMT}, and the response of the AugerPrime station electronics \cite{UUB}.
The following sections step through the construction of the detector in Geant4
and each stage of the simulation.
\begin{figure}[h]
\centering
\includegraphics[width=0.4\textwidth,clip]{{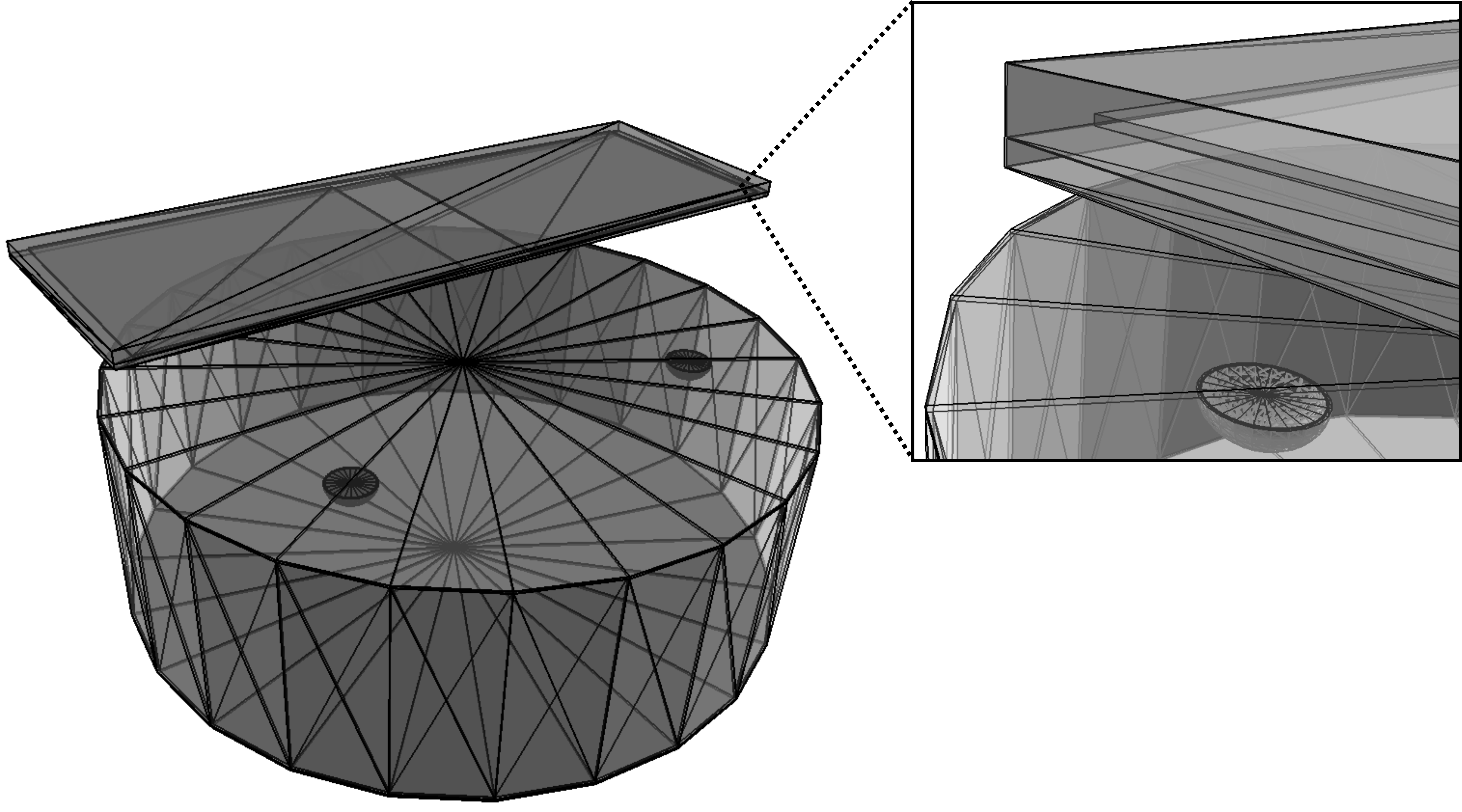}}
\caption{Visualization of both the \ac{SSD} and \ac{WCD} volumes defined in
the Geant4-based simulation application with a close-up of one corner of the
\ac{SSD}}
\label{f:g4_vis}
\end{figure}
\subsection{Detector Construction}
\label{s:detector}
\acp{SSD} are constructed within the same Geant4 world volume as their partner
WCDs. The volumes for both detectors may be observed in Fig. \ref{f:g4_vis}.
Regarding the \ac{SSD}, volumes for the aluminum casing, \ac{XPS}, \ac{EPS},
and scintillator bars themselves are defined according to the dimensions and
material properties of the production \ac{SSD} design, described in Sec.
\ref{s:ssd}. This is important to accurately reproduce the shielding above and
below the active area of the scintillator bars. Precise definition of the
position of the \ac{SSD} relative to the \ac{WCD} is important for preserving
signal correlations, as these depend on the fraction of particles that
intersect both detectors.
\begin{figure*}[t]
\def\figw{0.43}
\centering
\includegraphics[width=\figw\textwidth]{{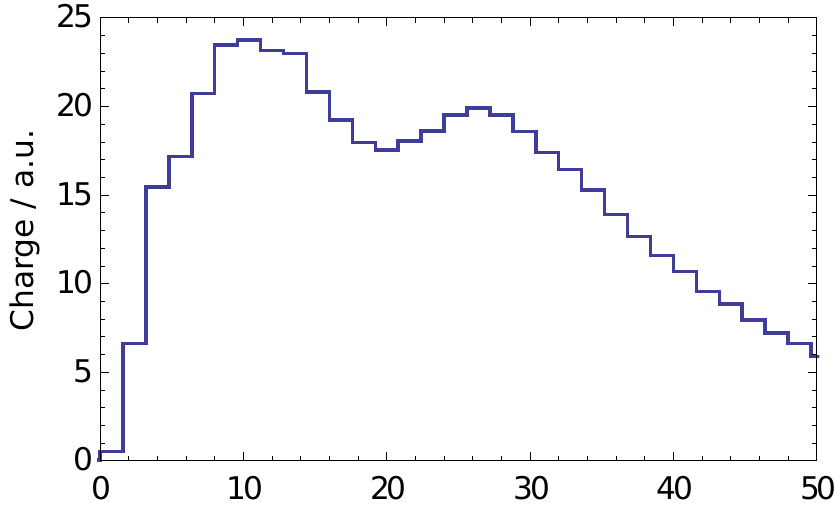}}
\includegraphics[width=\figw\textwidth]{{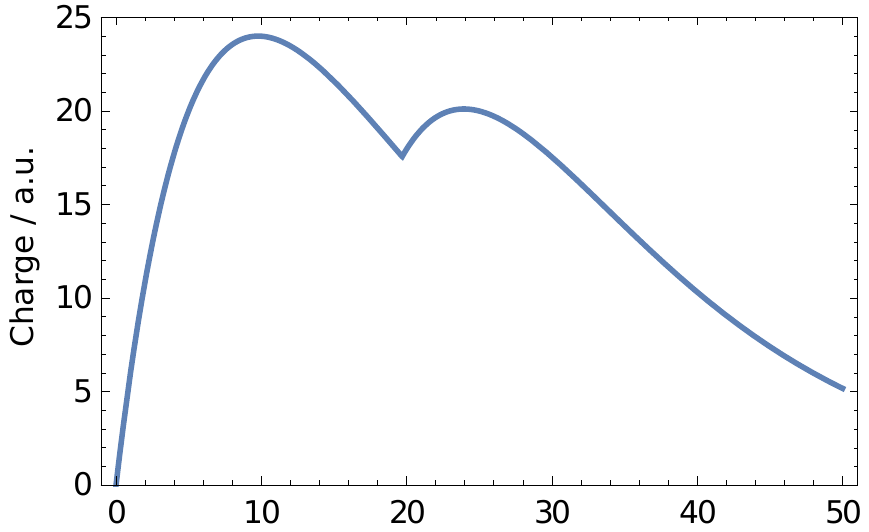}}
\\
\includegraphics[width=\figw\textwidth]{{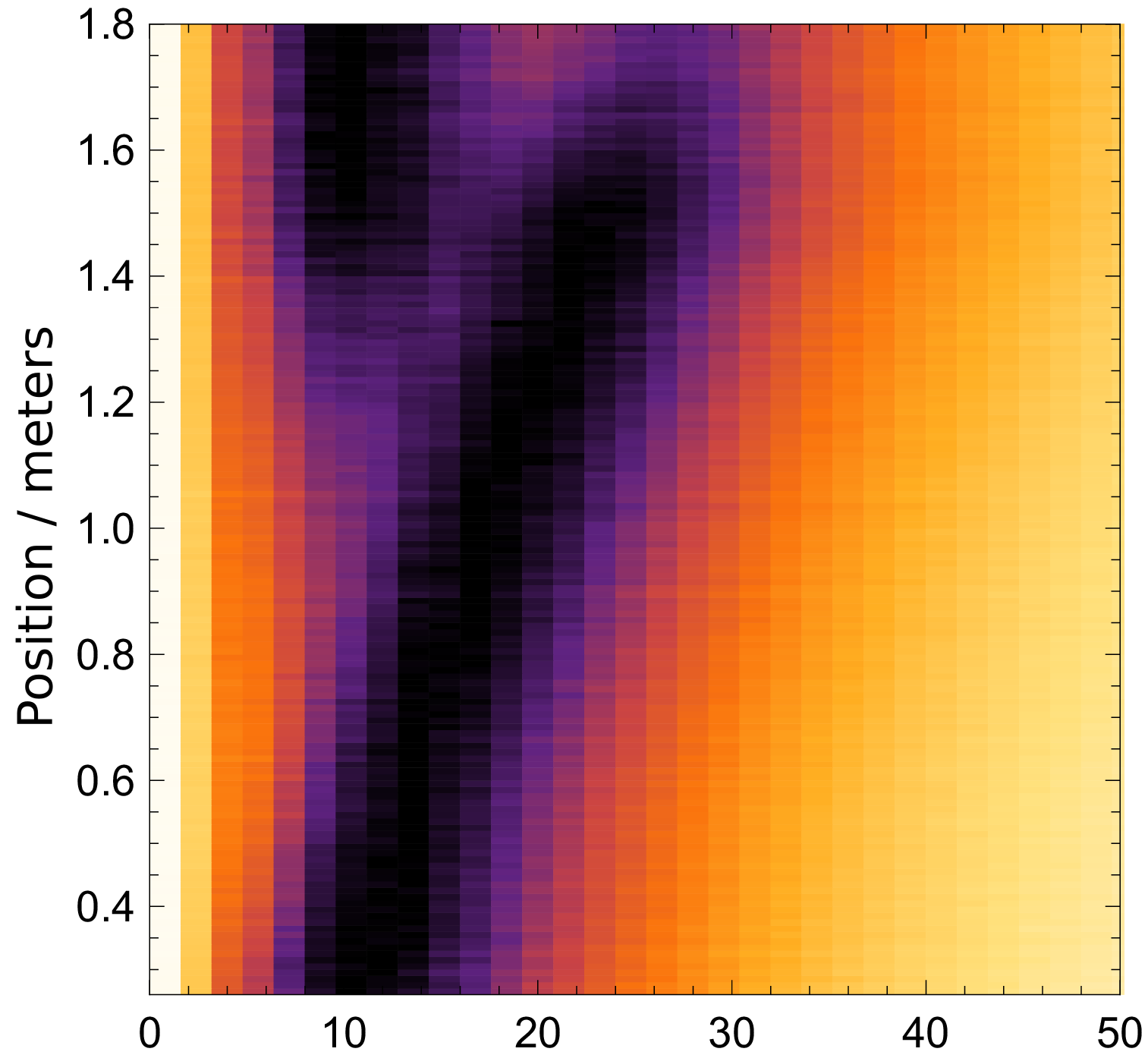}}
\includegraphics[width=\figw\textwidth]{{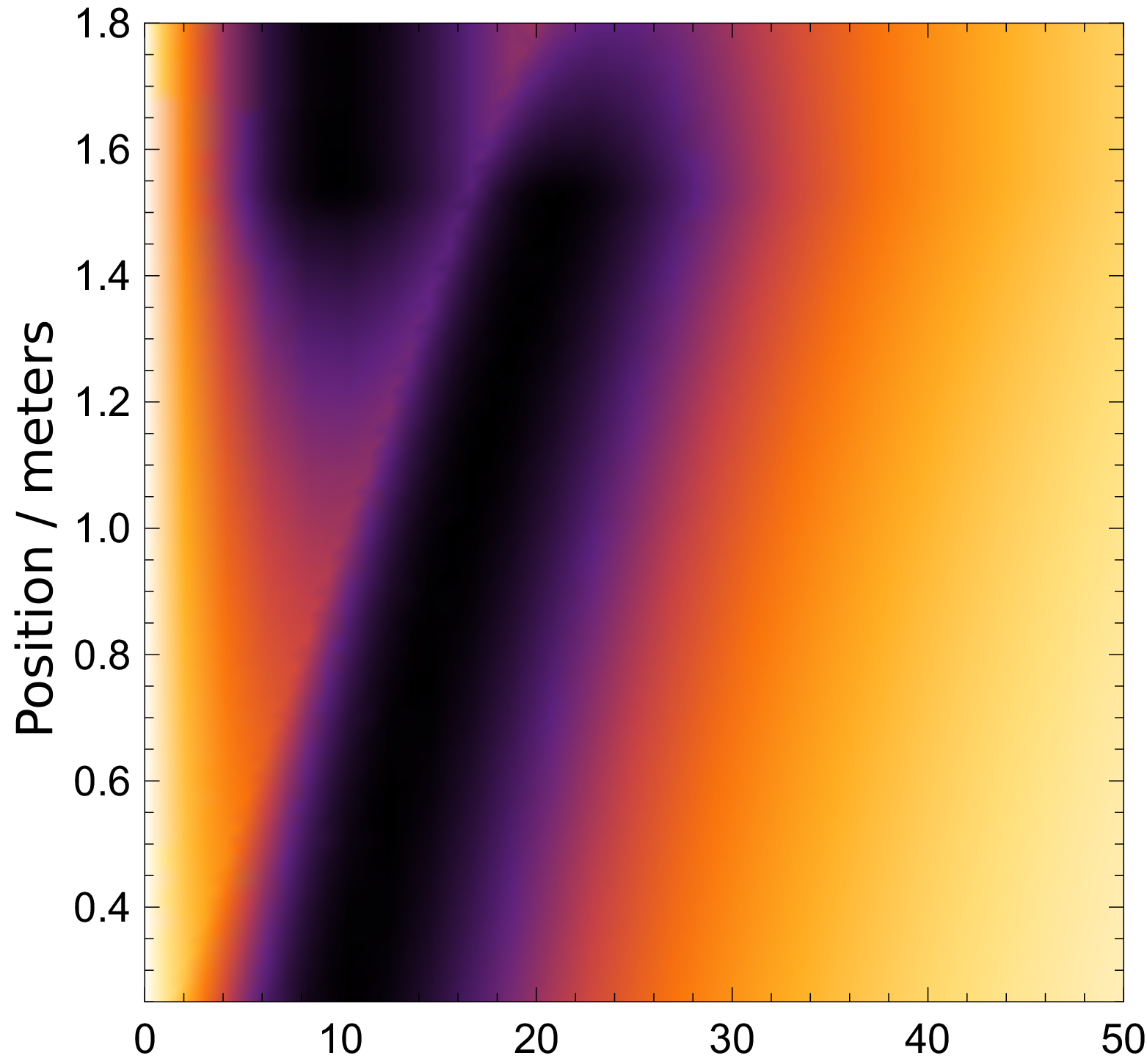}}
\\
\includegraphics[width=\figw\textwidth]{{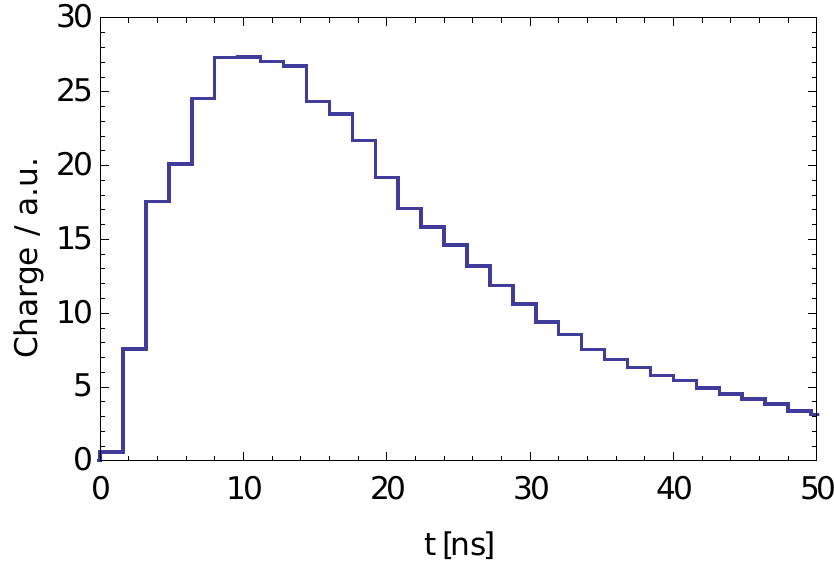}}
\includegraphics[width=\figw\textwidth]{{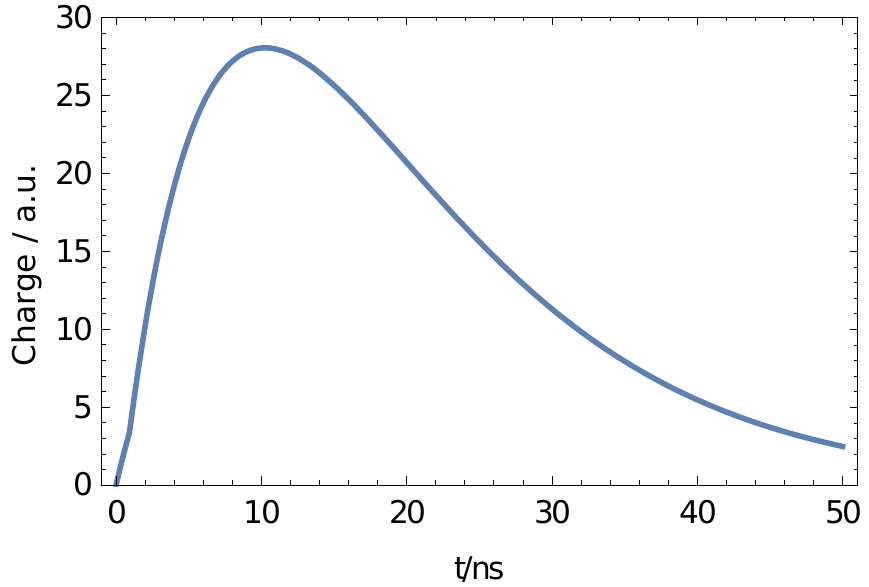}}
\caption{Depiction of the time evolution of the signal from single,
through-going muons in an \ac{SSD}. Muon telescope measurements are shown in
the plots on the left and the parameterization of these measurements are shown
in those on the right. \textit{Middle:} Density plot of the signal evolution
over one scintillator wing of the \ac{SSD}. The bottoms of the histograms
correspond to a particle crossing at the non-central edges of the scintillator
bars, where photons corresponding to the two legs of the signal traverse
approximately equal distance and therefore produce a unimodal distribution.
The tops of the histograms correspond to the central edges of the scintillator
bars, where the photons corresponding to the two legs of signal traverse
significantly different distances and therefore result in a bimodal
distribution. Slices in the horizontal axis have been normalized to the same
maximum. \textit{Top and bottom:} Signal shapes at the central and non-central
ends of the scintillator bars.}
\label{f:time_dists}
\end{figure*}
\subsection{\aclp{PE}}
\label{s:photoelectrons}
For each particle that intersects or is generated within the active
scintillator volumes, the energy deposited within the bars is extracted from
Geant4 and converted into a number of photoelectrons at the photo-cathode of
the \ac{PMT}. This conversion must take into account the attenuation of
photons within the bars and along the fibers as well as the excitations and
decays corresponding to scintillation and wavelength-shifting processes. The
conversion takes the form of
\begin{equation}\label{e:npe_ideal}
\bar{N}(\textbf{x}) = N_\mathrm{ref} \frac{E_\mathrm{dep}}{E_\mathrm{ref}}
f_\mathrm{att}(\textbf{x})
\end{equation}
where $E_\mathrm{dep}$ is the energy deposit in the scintillator bars,
$E_\mathrm{ref}$ and $N_\mathrm{ref}$ are the reference energy and \ac{PE}
number derived from simulations, and $f_\mathrm{att}(\textbf{x})$ describes
the degree of photon attenuation as a function of particle crossing position
$\textbf{x}$. $E_\mathrm{ref}$ is the peak in the distribution of energy
deposited in the \ac{SSD} by an ensemble of vertical, through-going muons
simulated for the volumes described in Sec. \ref{s:detector}. $N_\mathrm{ref}$
is a constant derived such that average value of Eqn. \ref{e:npe_ideal} when
integrated over all possible values of $\textbf{x}$ inside the scintillator
bars is equal to the the average number of \acp{PE} observed in real
measurements of the \acp{SSD} with the aforementioned muon telescope. For a
more detailed description of how these parameters were obtained, see
\cite{AugerPrimeOffline}. With this construction, the mean number of \acp{PE}
for a single, through-going muon is guaranteed to match real measurements. The
attenuation function $f_\mathrm{att}(\textbf{x})$ is written as
\begin{equation}\label{e:fatt}
f_\mathrm{att}(\textbf{x}) = A(\textbf{x}) L(\textbf{x}) \ ,
\end{equation}
where $A(\textbf{x})$ describes the attenuation along the \ac{WLS} fibers.
This attenuation is of the form
\begin{equation}
A(\textbf{x}) = e^{-\ell_\mathrm{c}(\textbf{x})/\lambda_\mathrm{f}} + e^{-\ell
\mathrm{f}(\textbf{x})/\lambda_\mathrm{f}} \ ,
\end{equation}
where the two exponential terms correspond to the two paths photons may travel
to reach the PMT due to the U-bend routing of the fibers. $\ell_\mathrm{c}$
($\ell_\mathrm{f}$) corresponds to the shorter (longer) travel path and is
calculated for a given crossing position $\textbf{x}$. $\lambda_\mathrm{f}$ is
the effective attenuation length of the fiber obtained through analysis of
muon telescope \ac{SSD} measurements. The term $L(\textbf{x})$ corresponds to
a decreased yield at crossing positions very close the edges of the
scintillator bars. Thus, for any energy deposit and particle crossing
position, an expected number of \acp{PE} corresponding to each of the
two paths photons may travel to reach the \ac{PMT} may be calculated. A
Poisson randomization is then performed on each of these expectations.
For each \ac{PE} coming out of these randomizations, two draws, one from each
of two exponential distributions are performed. This model of two exponential
decays, which may arbitrarily be attributed to the excitation associated with
scintillation in the scintillator bars and the wave-length shifting process in
the \ac{WLS} fibers, was found to provide a sufficient description of the data
\cite{AugerPrimeOffline}. The production time for each \ac{PE} is then
obtained via
\begin{equation}\label{e:petime}
t_\mathrm{PE} = t_\mathrm{global} + t_\mathrm{kin} + t_\mathrm{bar} +
t_\mathrm{fiber}
\end{equation}
where $t_\mathrm{global}$ is the global time obtained from Geant4,
$t_\mathrm{kin}$ is the kinematic delay of the photon as it traverses its
respective path to the \ac{PMT}, and $t_\mathrm{bar}$ and $t_\mathrm{fiber}$
are the times arising from the two aforementioned exponential decays.
The average measured and parameterized time distributions for crossing
positions along the complete active surface of one wing of the \ac{SSD}, which
compound Eqn. \ref{e:npe_ideal}, Eqn. \ref{e:petime}, and the convolution of
the two exponential decays, may be observed in Fig. \ref{f:time_dists}.
\begin{figure*}[t]
\centering
\includegraphics[height=0.345\textwidth,clip]
{{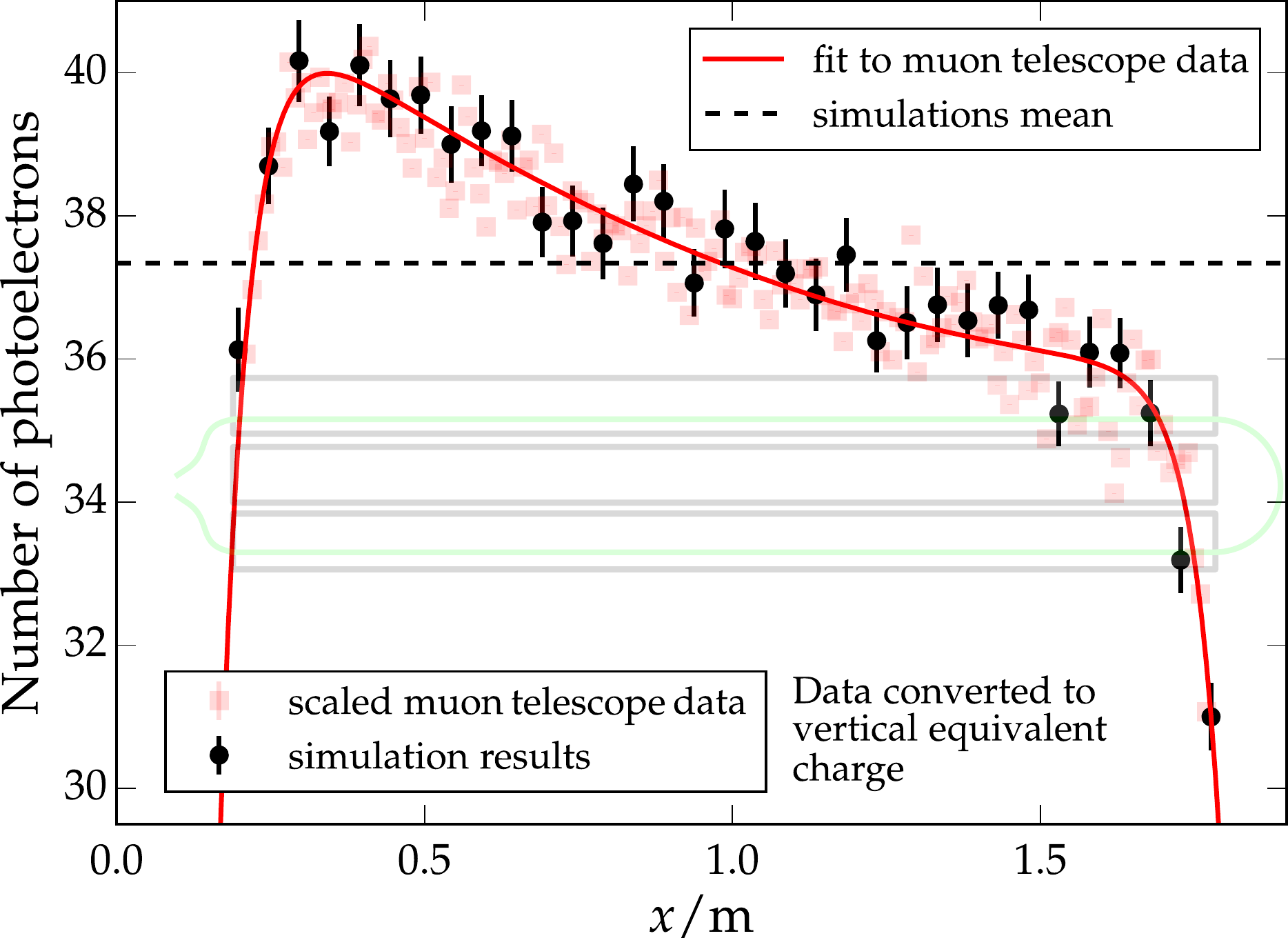}}\hspace{0.03\textwidth}
\includegraphics[height=0.355\textwidth,clip]{{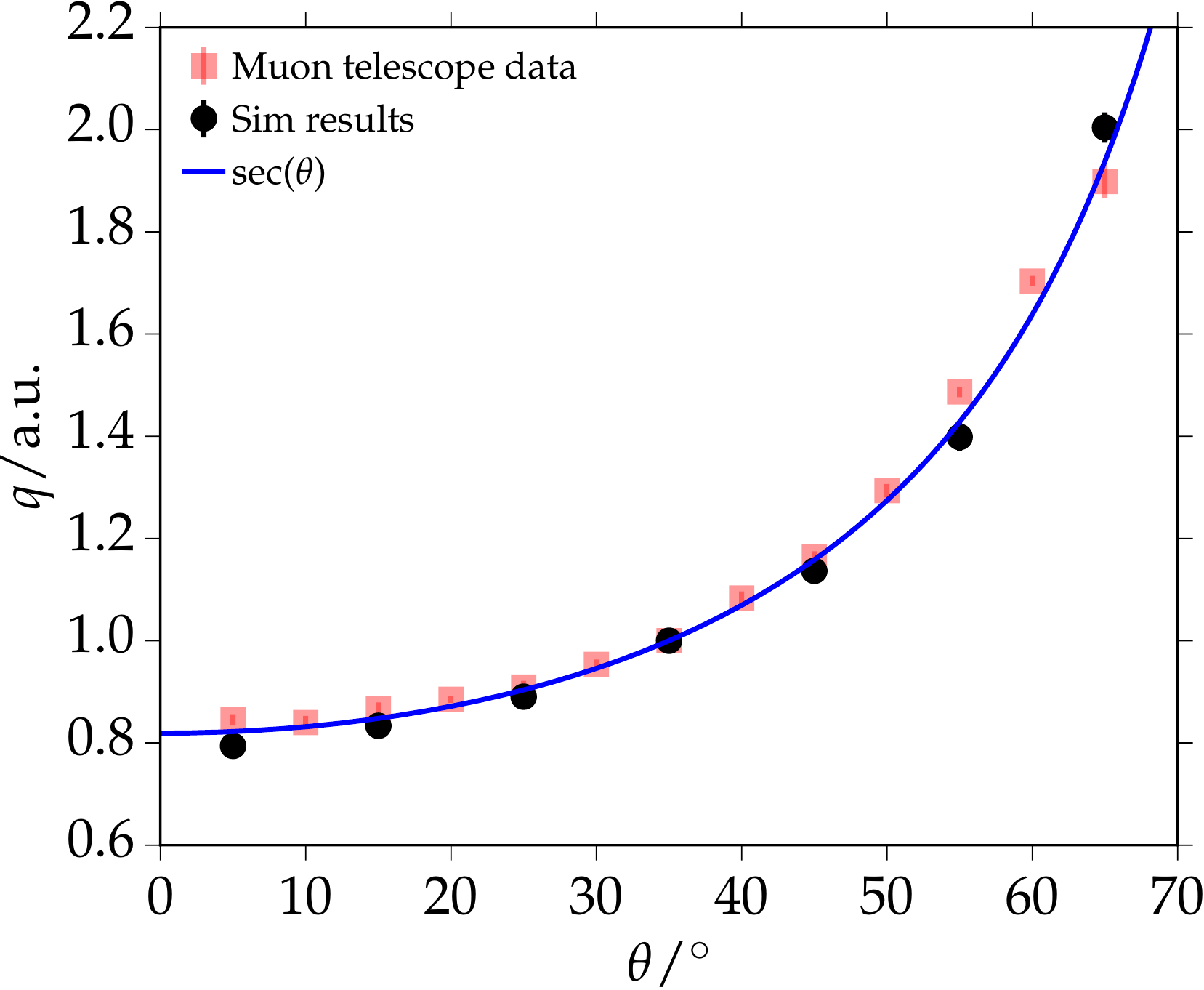}}
\caption{\textit{Left:} Average charge measured for single, through-going
muons as a function of the crossing position for one wing of an \ac{SSD}.
Measurements with the muon telescope are shown by the square, red markers,
whereas simulated data is represented by the circular, black markers. Larger
values on the horizontal axis correspond to crossing positions further from
the central edge of the scintillator, as depicted by the semi-transparent
schematic in the backdrop. \textit{Right:} Average charge as a function of
zenith angle. Once more, muon telescope measurements are given by the red
markers, whereas simulations are shown in black. Both data sets have been
scaled such that they equal one at an angle of \SI{35}{\degree} to compare
their evolution with tracklength ($1/\mathrm{cos}\theta$).}
\label{f:charge_validation}
\end{figure*}
\subsection{\acs{PMT} and electronics}
\label{s:pmt}
With a trace of PE production times in hand, the \ac{PMT} is simulated as
follows. For each \ac{PE}, the average pulse shape at the \ac{PMT} base
corresponding to an \ac{SPE} is scaled by a random draw from the corresponding
charge distribution. Both this pulse shape and the charge distribution were
measured for \ac{SPE} events for the \ac{PMT} in question, as described in
\cite{AugerPrimeOffline}. Performing this process for each \ac{PE} results in
a time distribution of the electrical current at the base of the \ac{PMT},
which is the input to the upgraded \ac{SD} station electronics \cite{UUB}.

The effect of the electronics on the time trace is simulated by convolving the
base current distribution with the measured electronics transfer function.
This results in an analog voltage trace, which is then scaled, sampled, and
digitized. The scaling is performed such that the resulting peak in the charge
distribution for a simulated ensemble of through-going muons equals that which
has been measured by prototype \acp{SSD} deployed in an \ac{EA} \cite{EA} at
the observatory site near Malargüe since 2016. The sampling and digitization
are performed in accordance with the \SI{120}{\mega\hertz} 12 bit design of
the AugerPrime electronics.
\section{Validation}
\label{s:results}
Validation of the output of the simulations was performed on a per-particle
basis by comparing the position-dependent charge and time distributions of
simulated muons to those measured by the muon telescope. In Fig.
\ref{f:charge_validation}, the dependencies of charge on crossing position and
zenith angle of the particles are compared between measurements and
simulations. In both cases, the simulations provide a reasonable description
of the data. A full validation closely examining the shapes of the charge and
time distributions is currently underway as are additional comparisons with
measurements of \acp{SSD} deployed in the \ac{EA}.
\begin{figure}[h]
\centering
\includegraphics[width=0.5\textwidth,clip]{{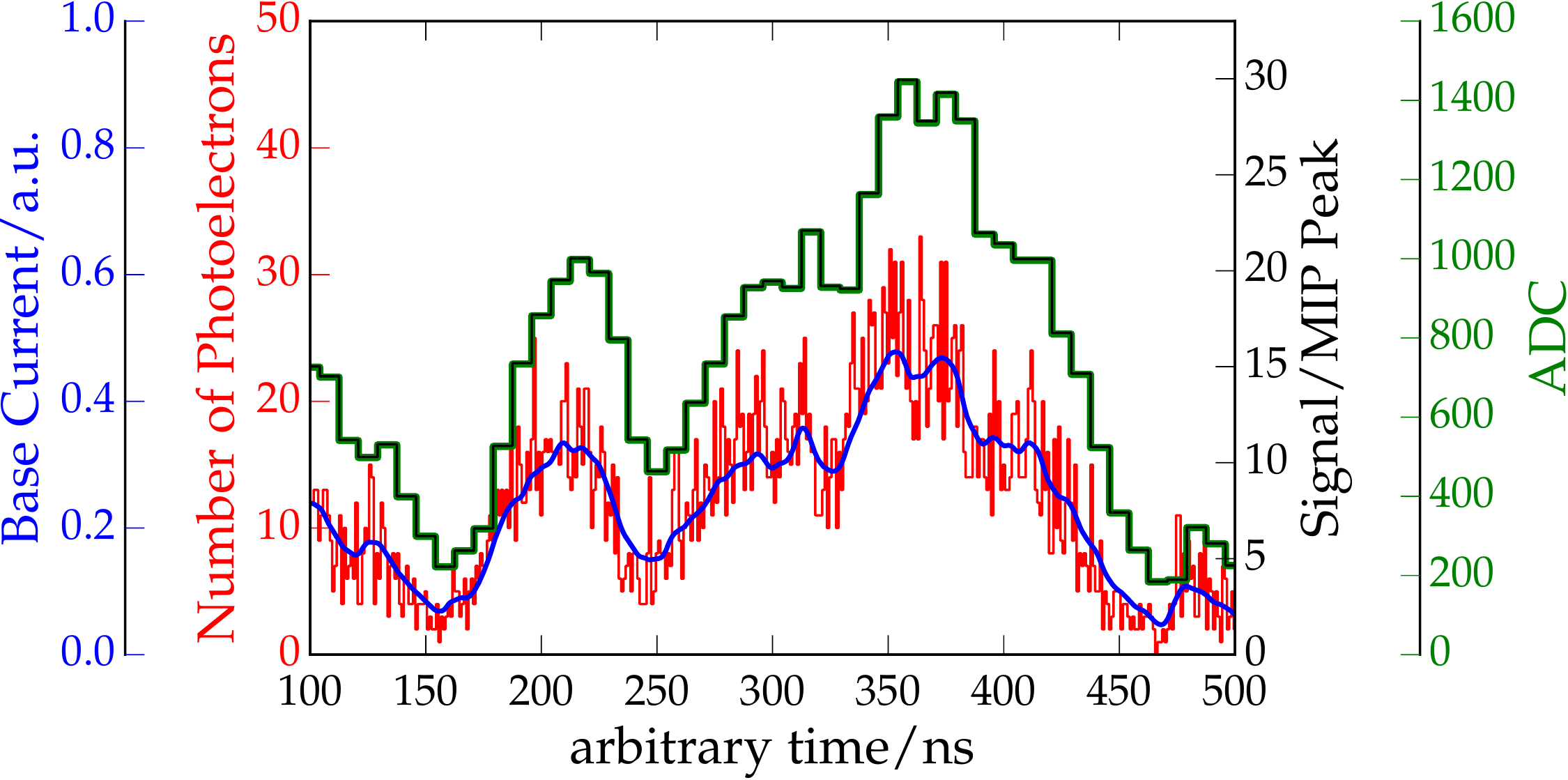}}
\caption{Simulated \ac{SSD} measurement. Time traces for \ac{PE} production,
current at the base of the PMT, post-electronics digitized signals, and
calibrated to the signal of one \ac{VMIP} are shown.}
\label{f:traces}
\end{figure}
\section{Example Application}
\label{s:application}
\begin{figure}[h]
\centering
\includegraphics[width=0.45\textwidth,clip]{{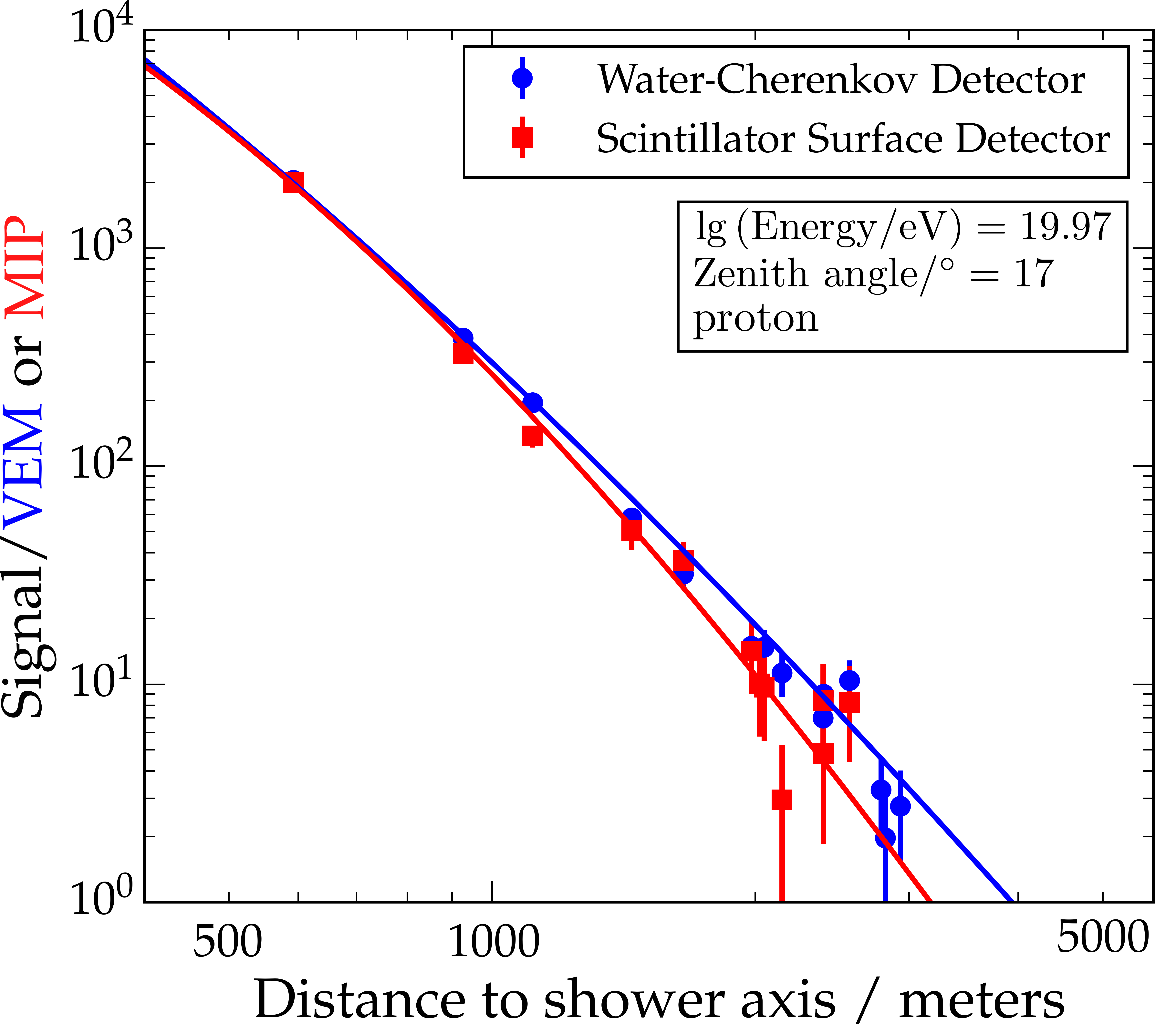}}
\caption{Simulated signals of \acp{SSD} alongside partner \acp{WCD} for
various samples of the lateral distribution of the particles reaching ground
in a simulated \ac{EAS}.}
\label{f:ldf}
\end{figure}

For a given simulated extensive air shower, the particles reaching ground near
each \ac{SD} station are injected into a cylindrical volume that completely
encases both the \ac{SSD} and \ac{WCD}. Propagation of these incident
particles into the shared volume and subsequent interactions and generation of
new particles is then left to the Geant4 software, and signals are simulated
as described in Sec. \ref{s:simulations}. A depiction of time distributions of
photoelectrons, current at the base of the PMT, and digital signals after
electronics processing are shown in Fig. \ref{f:traces} for an example station
simulation. An example event, where \ac{SSD} and \ac{WCD} pairs sample the
lateral distribution of particles at ground at various locations, based on
\ac{Auger}'s hexagonal grid, is depicted in Fig. \ref{f:ldf}.
\section{Conclusions}
\label{s:conclusions}
Detector simulations for the \ac{SSD} of the AugerPrime upgrade have been
developed and the methods, making use of Geant4 and real detector
measurements, have been described here. Initial validation has shown general
agreement between the detector simulations and centimeter-precision muon
telescope measurements for the charge and time distributions produced by
single, through-going muons. Additional, more detailed, validation is
currently being performed. The combined \ac{SSD} and \ac{WCD} simulation
application, housed in the \Offline detector simulation and reconstruction
software framework of Auger, has already and will continue to aid in the
interpretation of measurements performed with AugerPrime as well as in the
development and improvement of event reconstruction algorithms which include
primary mass. 
\end{document}